\documentclass[11pt]{article}

\usepackage{graphicx}
\usepackage{times}

\usepackage{amssymb}
\usepackage{mathtools}
\usepackage{mathrsfs}
\usepackage{gensymb}
\usepackage[utf8]{inputenc}
\usepackage{bm,color}
\usepackage{chngcntr}
\usepackage{setspace}
\counterwithout{figure}{section}
\counterwithout{figure}{subsection}
\counterwithout{figure}{subsubsection}

\usepackage{hyperref}
\hypersetup{
  colorlinks=true,        % false: boxed links; true: colored links
  linkcolor=blue,         % color of internal links
  citecolor=cyan,         % color of links to bibliography
}

\newcommand{\be}{\begin{equation}}
\newcommand{\ee}{\end{equation}}
\newcommand{\bes}{\begin{subequations}}
\newcommand{\ees}{\end{subequations}}
\newcommand{\bea}{\begin{eqnarray}}
\newcommand{\eea}{\end{eqnarray}}

\usepackage[width=170mm,headheight=110pt,top=25mm,bottom=25mm]{geometry}

\usepackage[numbers,sort&compress]{natbib}

\begin{document}

\begin{center}
{\LARGE PROBING THE EQUATION OF STATE OF NEUTRON STAR\\\vspace*{0mm} MATTER WITH GRAVITATIONAL WAVES FROM BINARY\\\vspace*{3mm} INSPIRALS IN LIGHT OF GW170817: A BRIEF REVIEW}
\end{center}
\vspace{0.1cm}

{\begin{center}
{Andreas Guerra Chaves$^{1,2}$~~and~~Tanja Hinderer $^{1,3}$}
\end{center}}

\vspace{0.1cm}

{$^1$ \emph{ GRAPPA, Anton Pannekoek Institute for Astronomy and Institute of High-Energy Physics, University of Amsterdam, Science Park 904, 1098 XH Amsterdam, The Netherlands}}

{$^2$ \emph{Maastricht Science Programme, Faculty of Science and Engineering, Maastricht University, P.O. Box 616 6200 MD Maastricht, The Netherlands}}

{$^3$ \emph{Delta Institute for Theoretical Physics, Science Park 904, 1090 GL Amsterdam, The Netherlands}}

\vspace{0.3cm}

\begin{abstract}

Neutron stars (NSs) are unique testbeds for exploring the physics of strongly interacting matter in extreme regimes of density, temperature, and isospin that are not accessible anywhere else in the Universe. The nature of neutron star matter can now be probed with gravitational-waves (GWs) from binary driven by nonlinear gravity, where phenomena such as tidal effects lead to characteristic matter-dependent GW signatures. We focus here on the dominant tidal GW imprints, and review the role of the characteristic tidal deformability parameter, its definition, computation, and relation to the equation of state. We briefly discuss the implications of the event GW170817, which enabled the first-ever constraints on tidal deformability from GW data. Finally, we outline  opportunities and challenges for probing subatomic physics with GWs, as the measurements will become more precise and will probe a diversity of the NS binary population in the coming years.
\end{abstract}
%------------------------------------------------------------------------
\section{Introduction}
\label{sec:Introduction}
\begin{figure}[h]
\label{fig:binarysetup}
\centering
\includegraphics[width=.7\textwidth]{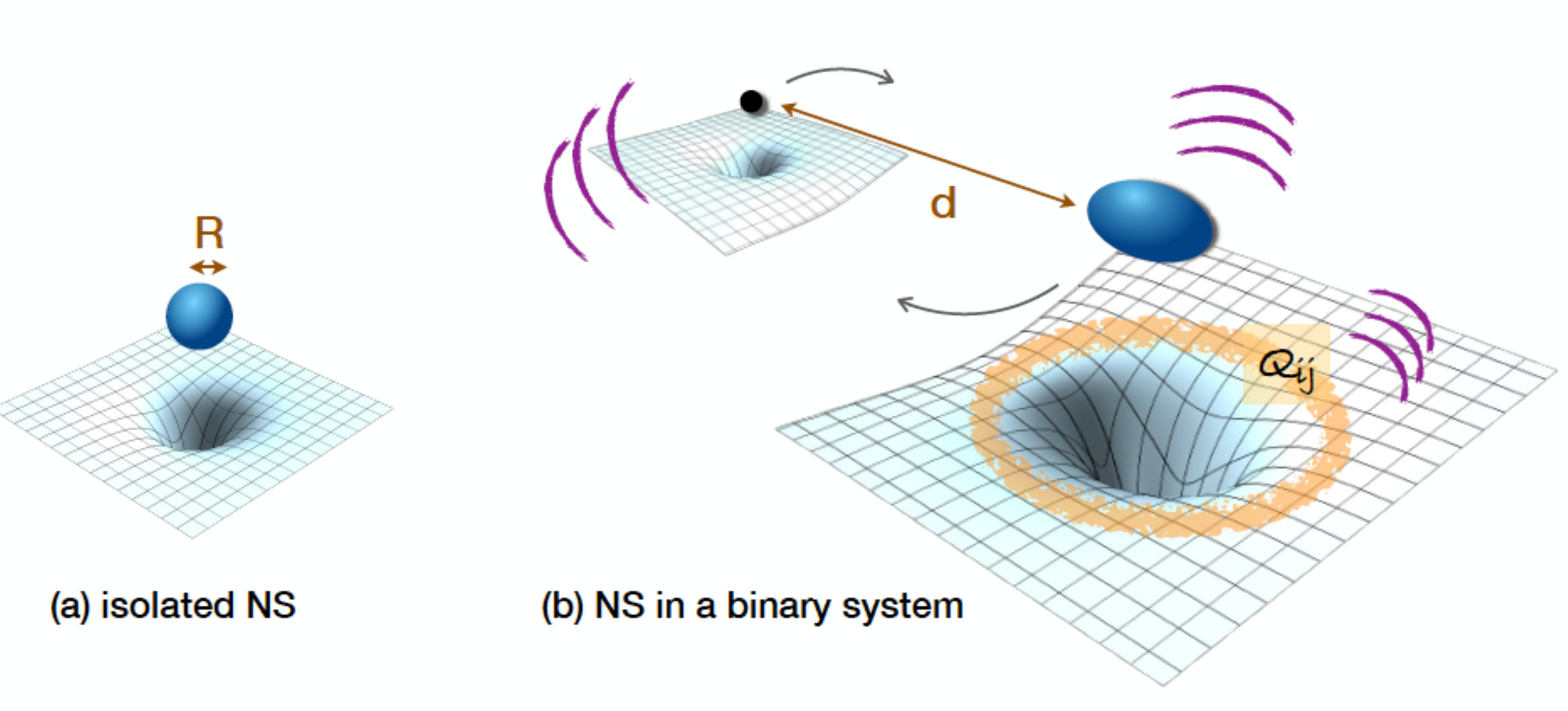}
\caption{Cartoon illustration of tidal effects in a binary system. \emph{Left panel}: An isolated non-spinning neutron star's mass distribution (blue sphere) and exterior spacetime (grid illustrating an equatorial slice through the spacetime, with the depth indicating curvature) are spherically symmetric. \emph{Right panel}: In a binary, the NS's matter distribution adjusts in response to the companion's spacetime curvature (or tidal fields). This distortion also manifests itself as structural changes in the NS's exterior spacetime geometry, which is the important quantity for GWs. The dominant tidal GW signature is characterized by the EoS-dependent tidal deformability parameter $\lambda=-Q_{ij}/{\cal E}_{ij}$, where $Q_{ij}$ is the induced quadrupole moment and ${\cal E}_{ij}$ characterizes the tidal field from the spacetime curvature sourced by the distant companion [see text]. These quantities are defined in the region of spacetime marked by the orange circle, at asymptotically large distances from the NS \emph{and} from the companion. Purple curves in the cartoon indicate dynamical distortions in the spacetime, which propagate nonlinearly and in part become the GWs measured by a distant detector.}
\end{figure}

\begin{figure}[h]
\centering
    \includegraphics[width=.8\textwidth]{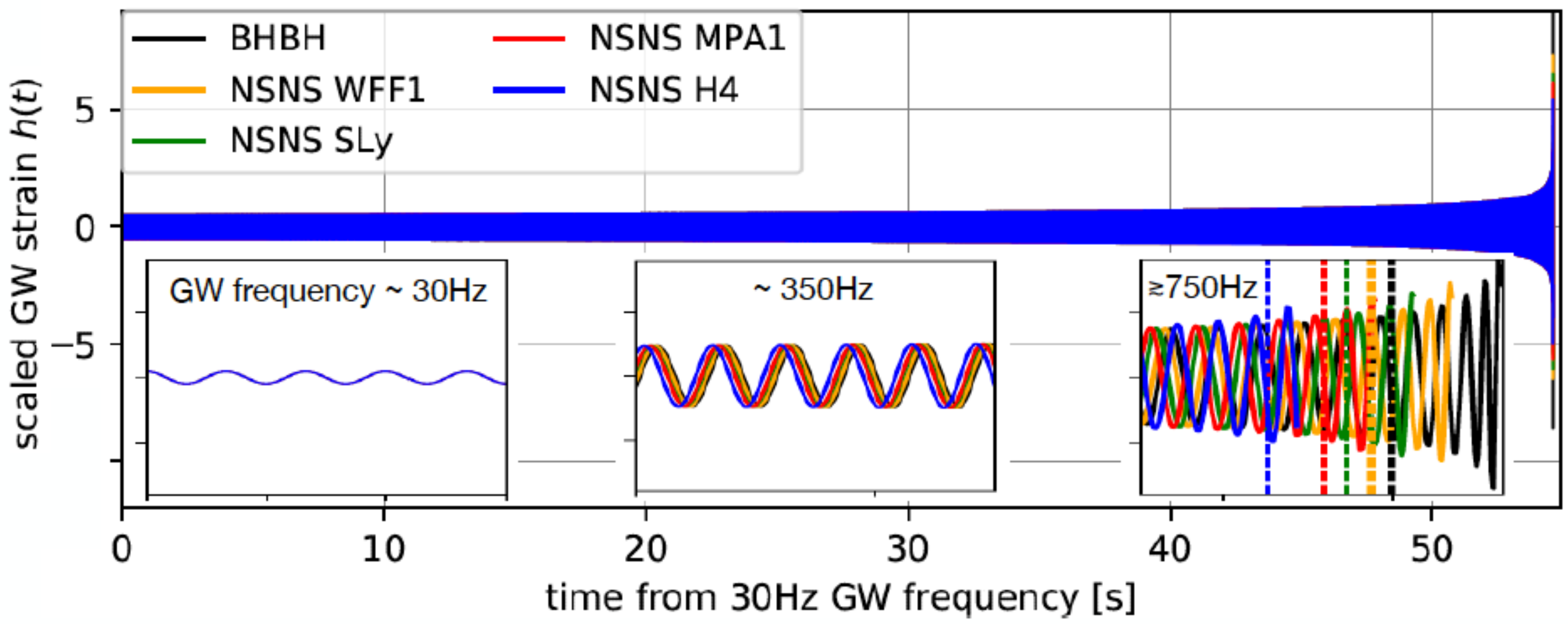}[h]
    \caption{\emph{Differences in GW signals due to tidal effects for different EoSs of NS matter} in nonspinning binary inspirals of two $1.4M_\odot$ NSs computed from the SEOBNRv4T waveform model~\cite{Hinderer:2016eia, Steinhoff:2016rfi, Bohe:2016gbl,Barausse:2009xi, Barausse:2011ys,Taracchini:2013rva}. The black curve corresponds to a point-mass (or black hole binary inspiral) waveform where matter effects are completely absent, while curves with different colors correspond to different EoSs for the NSs. The plot shows an inspiral where the waveforms for different systems all started out aligned, i.e. indistinguishable over a short time interval near GW frequencies of 30Hz after choosing the arbitrary time and phase shifts in the GW signals appropriately. The three inset plots show details at different stages of the evolution corresponding to GW frequencies around 30Hz (left), 350Hz (middle), and beyond 730-850 Hz (right), with the beginning of the range depending on the parameters and bounded by $\sim 730$Hz for the black hole binary and $\sim 850$Hz for the stiffest EoS example (H4). Near 30Hz, the waveforms are indistinguishable and dominated by the point-mass dynamics. After further inspiral, at GW frequencies of a few hundred Hz, small differences especially in the phase start to be discernible (middle inset). Tidal effects become largest in the late inspiral (right inset). Dashed vertical lines indicate that the configuration reaches a GW frequency of $1$kHz, beyond which current detectors become less sensitive. In this plot the waveforms terminate at the end of the binary's inspiral epoch, specifically at the parameter-dependent frequency of the peak GW amplitude predicted by fits to numerical relativity simulations of binary NSs~\cite{Dietrich:2017aum} and black holes~\cite{Bohe:2016gbl}. The postmerger signals are not shown. The axis label scaled strain refers to the overall scalings of $h$ with the orientation of the binary, its position in the sky, distance to the observer, and component masses; see e.g.~\cite{Cutler:1994ys} for the exact dependencies.} 
    \label{fig:waveforms}
\end{figure}

%------------------------------------------------------------------------
The gravitational-wave (GW) driven collisions of two neutron stars (NSs) at close to the speed of light offer unique opportunities for probing subatomic physics in unexplored regimes, at low temperature, nonvanishing isospin, and densities up to several times the normal nuclear density. Analogous to heavy ion collisions, the most prominent signatures of NS matter arise after the merger, where additional important information is contained in the associated electromagnetic counterparts and neutrino emission. However, the merger of two NSs generally occurs at GW frequencies $>1$kHz where the sensitivity of current detectors deteriorates, making it challenging to probe the rich NS physics beyond the inspiral with existing facilities ( LIGO~\cite{TheLIGOScientific:2014jea}, Virgo~\cite{TheVirgo:2014hva}, and  KAGRA~\cite{Aso:2013eba} due to come online soon). Nevertheless, GWs offer further intriguing possibilities to learn about properties of NS matter from the inspiral regime, before the merger. In this relatively clean epoch, phenomena such as tidal effects, illustrated in Fig.~\ref{fig:binarysetup}, imprint small changes in the GW signals that depend on the equation of state (EoS) of NS matter, as shown in Fig.~\ref{fig:waveforms}. A main characteristic EoS-dependent parameter influencing the GWs is the NS's tidal deformability, the ratio of the induced tidal deformation to the strength of the tidal perturbation due to the binary companion~\cite{Flanagan:2007ix}. Although the tidal GW signatures are small, the GW data analysis is very sensitive to such cumulative effects during the inspiral, especially to details of the phase evolution, as the measurements are based on cross-correlating the detector outputs with theoretically predicted signals to determine the best-matching parameters~\cite{Cutler:1994ys,Finn:1992xs, Cutler:1992tc}. The remarkable GW discovery of the NS binary GW170817~\cite{TheLIGOScientific:2017qsa,Abbott:2018wiz,Abbott:2018exr} enabled for the first time constraining the tidal deformability parameters from the data.  

In this review, we will focus on a basic description of the GW signatures of the dominant leading-order tidal effect and the associated characteristic EoS-dependent parameter. We aim to keep the descriptions as non-technical as possible, although some concepts from General Relativity (GR) will be required to understand the specific connection between GW signatures and fundamental NS physics. 
The organization of this review is as follows. We start with a qualitative overview of the methods used to model the binary system and the GW emission, outlining the basic ideas of how information about NS interiors propagates to become encoded in asymptotic GW signatures. We then introduce the tidal deformability parameter and discuss its computation and relation to the EoS. Next, we sketch the calculation of the associated leading-order GW imprint and the implications from the measurements of GW170817. We conclude with an outlook on the exciting prospects for probing NS matter with upcoming GW observations of a diversity of binary NSs and NSs with a black hole companion anticipated in the near future. 

%%%%%%%%%%%%%%%%%%%%%%%%%%%%%%%%%%%%%%%%%%%%%%%%%%
\section{Framework for describing matter effects during a binary inspiral}
\label{sec:theory}
%%%%%%%%%%%%%%%%%%%%%%%%%%%%%%%%%%%%%%%%%%%%%%%%%%
We consider a binary system of a NS with a distant companion as depicted in Fig.~\ref{fig:binarysetup}. A theoretical description to trace the connection between the microphysics of NS matter and the GW signatures requires using a tapestry of approximation schemes each amenable in different regions~\cite{Flanagan:1997fn,Racine:2004hs,Vines:2010ca,Vines:2011ud,Flanagan:2007ix,Blanchet:2006zz}:
\begin{itemize}
\item \emph{Body zone} (interior of the orange circle in Fig.~\ref{fig:binarysetup}): In the vicinity of the NS, gravity is strong as measured by the dimensionless self-gravity parameter $GM/Rc^2=O(1)$, where $M$ and $R$ are the NS's mass and radius. Describing the NS and its close proximity therefore requires full GR. The approximation is that from the perspective of the NS, the effect of the distant companion is reduced to the presence of an external tidal field, a gradient in gravity across the NS. The NS's response to this tidal perturbation can be calculated by considering linearized deviations from an isolated equilibrium configuration in GR~\cite{Hinderer:2007mb,Damour:2009vw,Binnington:2009bb}.   
\item \emph{Interaction zone} (exterior of the orange circle in Fig.~\ref{fig:binarysetup}): The orbital dynamics of the binary at large separation can be computed in the post-Newtonian approximation, where the mutual gravitational interaction and orbital velocity are treated as small parameters $\varepsilon_{\rm orbit}\sim GM/dc^2\sim v^2/c^2\ll 1$, with $d$ and $v$ being the orbital separation and velocity. This post-Newtonian description for point masses must be combined with an expansion in finite size effects, where the dimensionless expansion parameter is the ratio $\alpha=R/d\ll 1$. This yields a double expansion scheme~\cite{Flanagan:1997fn} in both the post-Newtonian parameter $\varepsilon_{\rm orbit}$ and the tidal expansion parameter $\alpha$. From the perspective of interaction-zone computations, the spacetime dynamics of the NS and its surrounding body zone are effectively reduced to those of its center-of-mass augmented with multipole moments, with the values of the multipole moments encoding the information from within the body zone. 
\item
\emph{Radiation zone}: The GWs are defined and measured in a region far away from the binary, where they manifest themselves as small distortions of spacetime away from flat space and propagate with the speed of light. As explained in detail in the review~\cite{Blanchet:2006zz}, the properties of the GWs are nonlinearly related to the multipole moments of their source, the binary system, which has contributions from both the orbital dynamics and the tidally induced multipoles.  
\end{itemize}

The connection between these three zones can be rigorously established using matched asymptotic expansions as described in Refs.~\cite{Flanagan:1997fn,Racine:2004hs,Vines:2010ca,Vines:2011ud,Blanchet:2006zz}; see also Refs.~\cite{Goldberger:2004jt,Steinhoff:2016rfi} for the same results within an effective action approach. The main information from the strong-field region impacting the dynamics and GWs thus proceeds through the asymptotic multipole moments of the body zone.

We will review in the next section how these multipole moments relate to the NS's internal structure, which will lead to the definition of the characteristic tidal deformability parameters encoding this information. We will omit any further details about the  approximation scheme mentioned above, and refer to the state-of-the-art results in Refs.~\cite{Vines:2011ud,Damour:2009wj,Vines:2010ca,Bini:2012gu, Steinhoff:2016rfi,Damour:2012yf}. The results of these perturbative calculations have also been used as inputs for improved descriptions such as effective one body (EOB) resummations that are used in GW data analysis~\cite{Buonanno:1998gg,Bohe:2016gbl,Babak:2016tgq,Nagar:2017jdw,Dietrich:2017feu,Steinhoff:2016rfi,Bernuzzi:2015rla, Hinderer:2016eia}. For further reading we also refer the reader to the review articles in Refs.~\cite{Yagi:2013awa,Yagi:2016qmr, Hinderer:2018mrj,Yagi:2016bkt}.

%%%%%%%%%%%%%%%%%%%%%%%%%%%%%%%%%%%%%%%%%%%%%%%%%
\section{Tidally perturbed neutron stars}
%%%%%%%%%%%%%%%%%%%%%%%%%%%%%%%%%%%%%%%%%%%%%%%%%
\subsection{Preliminaries: description of an equilibrium NS configuration}
\label{sec:equilibrium}
%%%%%%%%%%%%%%%%%%%%%%%%%%%%%%%%%%%%%%%%%%%%%%%%%
A nonspinning isolated NS in equilibrium is spherically symmetric, and its exterior spacetime at large distances is indistinguishable from that of a point mass or a black hole. The spacetime can be described by a metric $g_{\mu\nu}$ which gives a distance rule $ds^2=g_{\mu\nu}dx^\mu dx^\nu$. Here, Greek letters run over four spacetime coordinates $\{c t,x^i\}$, where $x^i$ denotes a set of three spatial coordinates, and a sum over repeated indices is implied. Specifically, the spacetime
can be described as~\cite{Hartle:1967he}
\be
ds_{\rm isolated~NS}^2 = -e^{\nu(r)} c^2 dt^2 + e^{\gamma (r)} dr^2 + r^2 (d\theta^2 + \sin^2 \theta d\varphi^2). \label{eq:NSmetric}
\ee
We model the NS matter by a perfect fluid stress-energy tensor 
$T_{\mu \nu} = c^{-2}(\epsilon + p) u_{\mu} u_{\nu} + p g_{\mu\nu}$, 
where $p$ and $\epsilon$ are the pressure and energy density and $u^\mu$ is the four-velocity of the fluid. 
Substituting these quantities into the Einstein field equations $G_{\mu\nu}=8\pi (G/c^4)  T_{\mu\nu}$ and energy momentum conservation $\nabla_\nu T^{\mu\nu}=0$ yields the following structure equations~\cite{Oppenheimer:1939ne}:
\be
\label{backgroundTOV}
\frac{d m}{dr} = \frac{4 \pi}{c^2} r^2 \epsilon, \ \ \ \ \ \ \ \ 
\frac{d \nu}{dr} = \frac{2G}{c^2}\, \frac{4 \pi r^3 p/c^2 + m}{r(r-2Gm/c^2)},  \ \ \ \ \ \ \ \ \ \frac{dp}{dr} = -\frac{G}{c^2}\frac{(4\pi r^3 p/c^2 + m) (\epsilon + p)}{r(r-2Gm/c^2)} .
\ee
The function $m(r)$ reduces to the NS's constant gravitational mass $M$ outside the matter distribution. The metric function $\gamma$ is set by $e^{-\gamma (r)} =1-2Gm/rc^2$.
Solving equations~\eqref{backgroundTOV} requires supplying an EoS  that relates $p$ and $\epsilon$. The system~\eqref{backgroundTOV} is then numerically integrated from the center at a small starting radius $\Delta r$ with $\Delta r \to 0$, a central density $\epsilon_c$ and mass $ m_c=(4\pi/3c^2) \epsilon_c (\Delta r)^3$, to the surface $r=R$ which corresponds to a vanishing pressure $p(R) = 0$. For an alternative and numerically often more convenient formulation see e.g.~\cite{Lindblom:2013kra}.

%%%%%%%%%%%%%%%%%%%%%%%%%%%%%%%%%%%%%%%%%%%%%%
\subsection{Definition of tidal deformability}
\label{sec:definitionoflambda}
%%%%%%%%%%%%%%%%%%%%%%%%%%%%%%%%%%%%%%%%%%%%%%
In a binary system, the presence of a companion distorts the spacetime curvature, giving rise to tidal fields. The curvature is measured by the Riemann tensor $R_{\mu \alpha\nu \beta}$, which is computed from second derivatives of the metric. In the rest frame of the NS, the quadrupolar tidal field sourced by the companion that is relevant for our purposes here is given by~\cite{Thorne:1980ru,Thorne:1997kt} 
\be
{\cal E}_{ij}=R_{titj}, \label{eq:Eijdef}
\ee 
where only the companion's contribution to the curvature enters on the right hand side. In Newtonian gravity the expression~\eqref{eq:Eijdef} reduces to ${\cal E}_{ij}=-GM_2 (\partial^2/\partial x^i \partial x^j)d^{-1}$, where $M_2$ is the mass of the companion.

From the NS's perspective the companion is moving, generating a tidal field that varies on multiples of the orbital timescale. The NS responds to this tidal disturbance by adjusting its internal structure to a new equilibrium configuration. These internal changes of the matter distribution also impact the spacetime geometry outside the NS, as discernible through multipole moments at large distances. The multipole moments associated with the exterior spacetime (analogous  to the gravitational potential in Newtonian gravity) are the key observable quantities that can be measured either with test particle orbits as in the Cassini flyby of Saturn's moon Titan~\cite{2012AGUFM.P23F..02I} or in a comparable-mass binary system where they influence the dynamics and GW signals~\cite{Bildsten:1992my,1994ApJ...426..688R,Lai:1993ve,Lai:1993di,1994ApJ...426..688R,1977A&A....57..383Z,1970A&A.....4..452Z,1978ASSL...68.....K,Kochanek:1992wk,Hansen:2005qv,Mora:2003wt,Kokkotas:1995xe,Flanagan:2007ix,Ferrari:2011as,PhysRevD.45.1017, doi:10.1143/ptp/91.5.871}.

The spacetime multipole moments are defined in a vacuum region surrounding the NS, at asymptotically large distances from its center as depicted by the orange circle in Fig.~\ref{fig:binarysetup}. In the scenario considered here, the multipole moments can be read off by writing e.g. the time-time component of the metric in an asymptotically Cartesian coordinate system whose origin is at the NS's center of mass as~\cite{Thorne:1980ru,Thorne:1997kt}
\be \lim_{r\to \infty} \frac{1+g_{tt}}{2c^{-2}   }=\frac{GM}{r}+ \frac{(3n^i n^j-\delta^{ij})Q_{ij}}{2r^{3}}+{\cal O}(r^{-4}) - \frac{1}{2} n^i n^j{\cal{E}}_{ij}r^2 +{\cal O}(r^3). \label{eq:Qdef}
\ee
Here, $r=\sqrt{\delta_{ij} x^i x^j}$, the quantity $n^i=x^i/r$ denotes a unit vector, and $\delta_{ij}$ is the Kronecker delta. In this setting, the $\ell$th mass multipole moment is associated with the piece in the asymptotic expansion~\eqref{eq:Qdef} that scales with the distance from the NS as $r^{-(\ell+1)}$, although a small ambiguity remains in these definitions ~\cite{Thorne:1984mz}. Specifically, the coefficient of the monopole term $\propto 1/r $ is the gravitational mass $M$, the dipole term $\propto r^{-2}$ vanishes because the origin of the coordinates coincides with the NS's center-of-mass, and the coefficient of the quadrupole term $\propto 1/r^3$ defines the NS's mass quadrupole moment tensor $Q_{ij}$. In Newtonian gravity, $Q_{ij}$ can be expressed as an integral over the density perturbation, however, in GR $Q_{ij}$ must be computed from the Einstein field equations as we discuss next.

To calculate the relation between $Q_{ij}$ and the properties of NS matter we first note that the NS's response to the tidal perturbation can be described in terms of excitations of the NS's oscillation modes. The modes are characterized by a set of integers $(n, \ell, m)$, where $(\ell,m)$ are the angular quantum numbers associated with a spherical harmonic decomposition and $n$ corresponds to radial nodes. In a binary inspiral, the NS's modes are either resonantly excited when the tidal forcing frequency coincides with the mode frequency or adiabatically driven when the two frequencies differ significantly. The tidally induced quadrupole moment $Q_{ij}$ is a sum of contributions from all quadrupolar or $\ell=2$ oscillation modes of the NS, $Q_{ij}=\sum_{n}Q_{ij}^n$, where $n$ denotes the modes with different radial nodes. In the situation we are considering, this sum is dominated by the contribution from the NS's fundamental modes corresponding to $n=0$, which have by far the strongest tidal coupling~\cite{Kokkotas:1995xe}. For simplicity we will therefore specialize to the case of fundamental modes and omit the sum over all other modes that contribute to $Q_{ij}$ in the subsequent discussion.

It is useful to first consider the \emph{adiabatic} limit, where the NS's internal time scales associated with the fundamental modes $\tau^{\rm int}\sim \sqrt{R^3/GM}$ are fast compared to the time scale of variations in the tidal field $\tau_{\rm orb}\sim \sqrt{d^3/GM_{\rm T}}$ where $M_{\rm T}=M_1+M_2$ is the total mass of the binary. In this limit the induced quadrupole is linearly proportional to the tidal field through a constant response coefficient $\lambda$, the tidal deformability parameter~\cite{Flanagan:2007ix}:
\be
Q_{ij}^{\rm adiab}=-\lambda \, {\cal E}_{ij}.\label{eq:lamdef}
\ee
The tidal parameter $\lambda$ is related to the  tidal Love number~\cite{AEHLove} or apsidal constant  $k_2$ and NS radius by $\lambda=2/(3G) k_2 R^{5}$.
In many contexts it is useful to work with the dimensionless tidal deformability
\be
\Lambda=\frac{\lambda\, c^{10}}{G^4\, M^5}.
\ee
We next discuss the computation of $\lambda$ and its dependence on the NS's internal structure.

%------------------------------------------------------------------------
\subsection{Computation of tidal deformability}
\label{sec:LambdaGR}
%------------------------------------------------------------------------
Computing the tidal response coefficient $\lambda$ requires solving the Einstein field equations and stress-energy conservation for linear quadrupolar perturbations to the equilibrium configuration described in Sec.~\ref{sec:equilibrium}. For this calculation, it is most convenient to use a spherical coordinate system instead of the Cartesian coordinates used above. To convert all the Cartesian tensors to a spherical decomposition we start by writing out the unit vector components $n^i=(\sin\theta \cos\phi,\sin\theta \sin\phi, \cos\theta)$. From these expressions, we can compute an explicit transformation between unit vectors and spherical harmonics $Y_{\ell m}$. For example, the $(\ell,m)=(1,1)$ spherical harmonic $Y_{11}(\theta, \phi)=-\sqrt{3/8\pi}\sin\theta e^{i\phi}$ can be written as the linear combination of unit vector components $Y_{11}(\theta, \phi)={\cal Y}^{j}_{11}n_j$, where the summation on the repeated index $j$ is implied, and ${\cal Y}^j_{11}$ is an array of constants. We can read off the components of ${\cal Y}^j_{11}$ from the angular dependences of $Y_{11}(\theta, \phi)$ and $n^i$ to be ${\cal Y}^j_{11}=-\sqrt{3/8\pi}(1,i,0)$. Similarly, the quadrupolar harmonics with $\ell=2$ are related to bilinear products of unit vectors through $Y_{2m}(\theta, \phi)={\cal Y}^{ij}_{2m}\left(n^i n^j-\delta_{ij}/3\right)$, with the quantities ${\cal Y}^{ij}_{2 m}$ being matrices with constant coefficients. Their component can be explicitly worked out from substituting the explicit expressions for $Y_{2m}$ and $n^i$ or by using the general formula given e.g. in Ref.~\cite{Thorne:1980ru}, where we refer the reader for further details. 

These relations imply that a Cartesian tensor such as $Q_{ij}$ can be expanded into its spherical harmonic decomposition as $Q_{ij}=2/3 \sum_m Q_m {\cal Y}_{ij}^{*\, 2m}$, where the prefactor is due to the normalization and the fact that we are using the inverse transformation than discussed above. A similar expansion applies for the tidal tensor ${\cal E}_{ij}$. The spherical harmonic analogue of ~\eqref{eq:lamdef} is then given by $
Q_{m}^{\rm adiab}=-\lambda \, {\cal E}_{ m}$, and it is sufficient to consider a single fixed value of $m$, with $m=0$ being the simplest case. While we are mainly interested in the quadrupole $\ell=2$ here, we will keep the multipolar index $\ell$ general as it requires only minor changes.

We next use these considerations to express all the quantities appearing in the Einstein field equations (metric components, fluid variables) as the equilibrium background solution plus a linear perturbation that is decomposed into spherical harmonics. For instance, the time-time component of the perturbed metric, from which we can determine $Q_{ij}$ as explained in Eq.~\eqref{eq:Qdef}, can be written as $g_{tt}=-e^{\nu(r)}\left(1+\delta g_{tt}\right)$, where the prefactor is the metric function describing the equilibrium configuration and the perturbation can be decomposed as~\cite{PhysRev.108.1063}
\be
\delta g_{tt}= H(r)Y_{\ell m}(\theta, \varphi).
\label{eq:decompose}
\ee
In general, the perturbed quantities also depend on time, which is usually taken to be of the form $\sim e^{i\omega t}$, however, since we are working in the adiabatic limit of tidal effects, we have already specialized to static perturbations. A similar decomposition as for $\delta g_{tt}$ in Eq.~\eqref{eq:decompose} applies for the other independent metric components with different as yet undetermined functions of $r$, as well as for the perturbations to the matter quantities ($p$,$\epsilon$,$u^\alpha$). Substituting these decompositions into the Einstein field equations and stress-energy conservation, using the normalization of the four-velocity $g_{\alpha \beta}u^\alpha u^\beta=-1$, and linearizing in the perturbations leads to various relations between the radial functions characterizing the perturbations in the metric and the matter variables. This enables reducing the perturbed system to the following single differential equation for $H(r)$~\cite{Hinderer:2009ca,Hinderer:2007mb, Damour:2009va, Binnington:2009bb}:
\bea
\label{eq:H0}
0&=&\frac{d^2 H}{dr^2} + \left[ r + \frac{Gm}{c^2} e^{\gamma} + \frac{2 \pi Gr^3}{c^4} (p-\epsilon) e^{\gamma}\right] \frac{2}{r^2} \frac{dH}{dr} \nonumber\\
&+& \left\{ e^\gamma \left[ \frac{4 \pi G}{c^4} (\epsilon + p) \frac{d\epsilon}{dp} + \frac{4 \pi G}{c^4} (5 \epsilon + 9 p)- \frac{\ell (\ell +1)}{r^2} \right]  - \left( \frac{d\nu}{dr} \right)^2 \right\} H . \;
\eea
 The initial condition at the center, for $r\to 0$, is $H \propto (\Delta r)^\ell$ to ensure regularity of the solution. The constant of proportionality is irrelevant in further calculations of the tidal deformability and can be chosen arbitrarily.

Outside the NS, the metric perturbation reduces to the general form
\be
H =  c_1\,  {Q}_{\ell 2}(c^2r/GM -1)+c_2 \, {P}_{\ell 2}(c^2r/GM -1), \label{eq:exter}
\ee
where ${P}_{\ell 2}$ and ${Q}_{\ell 2}$ are the associated Legendre functions of the first and second kinds respectively. The constants $c_1$ and $c_2$ can be related to $\lambda$ by using the asymptotic expansion of the Legendre functions in \eqref{eq:exter} for large arguments, comparing with the definition of $Q_{ij}$ and ${\cal E}_{ij}$ in the asymptotic metric \eqref{eq:Qdef} and using the definition of $\lambda$ from \eqref{eq:lamdef}. Matching the interior and exterior solutions at the NS's surface and eliminating ${\cal E}_{ij}$ by considering the logarithmic derivative of \eqref{eq:exter} leads to the following explicit algebraic expression~\cite{Hinderer:2009ca,Hinderer:2007mb}

\begin{align}
\label{k2}
\Lambda & = \frac{16}{15} (1-2 C)^2[2+2C(y-1)-y]\nonumber\\
&\times \left\{2C[6-3y+3C(5y-8)]  +4C^3[13-11y+C(3y-2)+2C^2(1+y)] \right. \qquad\qquad\nonumber\\ &\left. \; \; \; \; +3(1-2C)^2[2-y+2C(y-1)] \ln(1-2C)  \right\}^{-1}, \qquad
\end{align}
where $C=GM/Rc^2$ is the NS's compactness and 
\be
\label{eq:y}
y\equiv \frac{r}{H} \frac{dH}{dr} \bigg\rvert_{r=R},
\ee
For an incompressible star with constant density (as applies e.g. for strange quark matter stars), the density profile is a step function at $r=R$ and the matching of the interior and exterior solutions for $H$ must take into account a correction from the discontinuity given by~\cite{Damour:2009vw}
\be
\label{correcty}
y^{\rm out}_{\rm incompressible} = y^{\rm in}- 3, 
\ee
where $y^{\rm in}$ is the solutions obtained from the numerical integration of \eqref{eq:H0} in the interior, and $y^{\rm out}$ is to be substituted into the expression \eqref{k2} for $\Lambda$.

The generalization to higher multipoles entails solving the differential equation~\eqref{eq:H0} with the appropriate value of $\ell$, and using the following expression for $\Lambda_\ell$~\cite{Damour:2009vw,Binnington:2009bb}:
\be
\label{eq:kl}
(2\ell-1)!!\Lambda_\ell = -\left.\frac{{P}_{\ell 2}'(z)-C \, y {P}_{\ell 2}(z)}
{{Q}_{\ell 2}'(z)- C\, y\, {Q}_{\ell 2}(z)}\right\vert_{z=1/C-1}, 
\ee
where $y$ is the quantity \eqref{eq:y} for $H$ computed from \eqref{eq:H0} with the appropriate value of $\ell$.

 To summarize, the computation of $\Lambda$ proceeds through the following steps:
(1.) obtain the background solution from \eqref{backgroundTOV}, 
(2.) numerically solve for the perturbations in the interior from \eqref{eq:H0} and evaluate the results at the NS's surface to compute \eqref{eq:y}, and (3.) substitute in Eq.~\eqref{k2}.

For the case of a nonspinning black hole, which involves only vacuum spacetime curvature without any matter, the tidal deformability vanishes $\lambda_{\rm black~hole}=0$~
\cite{Damour:2009va, Kol:2011vg, Chakrabarti:2013tca, Porto:2016zng}. 
This means that the asymptotic multipole moments of a black hole spacetime remain unaltered under tidal perturbations~\cite{Gurlebeck:2015xpa} even though the geometry of the horizon responds to the tidal disturbance by deforming away from spherical symmetry, see e.g.~\cite{Hartle:1974gy,OSullivan:2014ywd, Poisson:2009di} for further discussion and references. This behavior illustrates the nonlinear relationship between multipole moments of local surface deformations and of the asymptotic spacetime of an object at large distances in GR. By contrast, for objects described by Newtonian gravity such as the planets there is a simple, linear, universal relationship between the dimensionless response coefficients for multipoles characterizing the deformation of the matter distribution and the corresponding changes in the exterior gravitational field. 

%%%%%%%%%%%%%%%%%%%%%%%%%%%%%%%%%%%%%%%%%%%%
\subsection{Link to the Equation of State}
\label{sec:RelatEOS}
%%%%%%%%%%%%%%%%%%%%%%%%%%%%%%%%%%%%%%%%%%%%
As mentioned before, solving~\eqref{backgroundTOV} and~\eqref{eq:H0} requires specifying a relation between $\epsilon$ and $p$. This relation is given by the NS EoS and therefore determines the global physical properties of the NS. Matter in NS interiors is expected to reach 5-10 times the nuclear saturation density, $\rho_{sat}\simeq2.8\times10^{14}g/cm^3$, an uncharted density regime with potential for new physics. Such extreme conditions are not accessible to terrestrial collider experiments and are theoretically challenging. Hence, the EoS is better understood in the low-density regime, corresponding to the outer parts of the NS. However, the nature and composition of matter in the high-density inner cores of NSs, dominated by complex multibody interactions, remain largely unknown.
At first, NSs were hypothesized to contain mainly neutrons, and if treated as a Fermi Gas (i.e. only considering quantum degeneracy pressure), integration of equations~\eqref{backgroundTOV} would yield a maximum mass of 0.7$M_{\odot}$ \cite{Oppenheimer:1939ne}. Recent measurements of NS masses reaching about 2$M_{\odot}$  \cite{antoniadis_massive_2013, 2010Natur.467.1081D,Cromartie:2019kug}  evidence the utter importance of the particle interaction scheme employed and the constituents considered.

Accordingly, a broad range of theoretical models have been designed that differ in assumptions about the composition, multi-body interactions, symmetry energy, computational schemes, and various other inputs. Some EoSs involve only nucleonic matter, however, since nucleons start to geometrically overlap at $\sim 4\rho_{sat}$ \cite{Ozel:2016oaf, Baym:2017whm, Haensel:2007yy,2018ASSL..457.....R}, new forms of matter might appear at higher densities. For example, a gradual increase on the nucleon overlapping can lead to phase transitions to quark matter (e.g.,\cite{Kojo:2015nzn}) or to Bose-Einstein condensates of pions (e.g.,\cite{Akmal:1998cf}) or kaons (e.g.,\cite{Kaplan:1986yq}). 
 
These are only a small sample of the possibilities, and many other models have been proposed; see e.g. Ref.~\cite{Ozel:2016oaf, 2017hsn..book.1331H,Baym:2017whm, Haensel:2007yy,2018ASSL..457.....R} for a more comprehensive list.

Furthermore, the relation between the pressure and the energy density of the region depends not only on the kinds of constituents considered but is also significantly impacted by the choices for modeling the interaction between these particles. While two-body forces are expected to be dominant around $\rho_{sat}$, approaches involving
higher n-body interactions are needed for higher densities. For a detailed explanation of these schemes see Ref.~\cite{Baldo:2011gz,Gandolfi_review,Haensel:2007yy}. 

\begin{figure}
\centering
\includegraphics[width=1\textwidth]{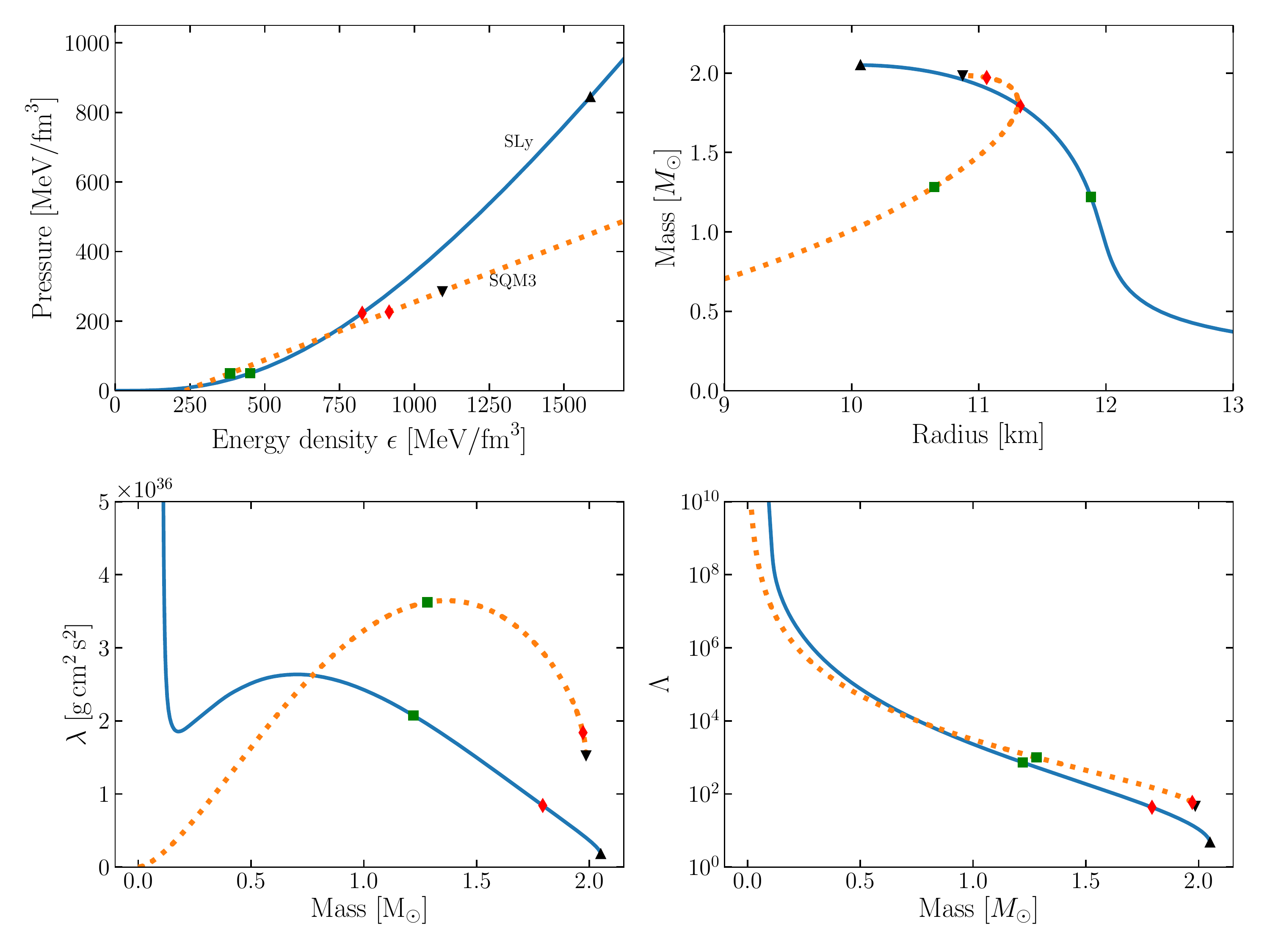}
\caption{\emph{Relation between the EoS and the macro-physical properties of a NS for the EoSs SLy (blue) and SQM3 (orange, dashed).} \emph{Top Left:} Microphysical relation between pressure and energy density as given by the two EoSs considered here. The black triangles represent the maximum central pressure a stable star can support as given by the EoS and the integration of Eqs.~\eqref{backgroundTOV} and \eqref{eq:H0}. The red diamonds and the green squares represent equal central pressures for start with the two distinct EoSs. These then translates into NSs with very distinctive characteristics, as illustrated by the other panels. Here we use units with $c=1$. \emph{Top Right:} Mass to radius relation, where a higher central pressure results in larger mass for both EoSs, and smaller radius for the nucleonic EoS (SLy) and generally higher radii for the incompressible EoS (SQM3). \emph{Bottom left:} Tidal deformability versus mass. Similar to the trends in the mass-radius relation, smaller central pressures give larger deformabilities for standard compressible NSs. Again, the incompressible SQM3 EoS yields a distinct behavior. \emph{Bottom right:} Dimensionless tidal deformability plotted as a function of mass, where smaller masses also yield higher deformabilities but details are less apparent. The dimensionless nature of this parameter hides the exotic behaviour of SQM3.}
\label{fig:4panelSLY}
\end{figure}

\begin{figure}
\centering
\includegraphics[width=1\textwidth]{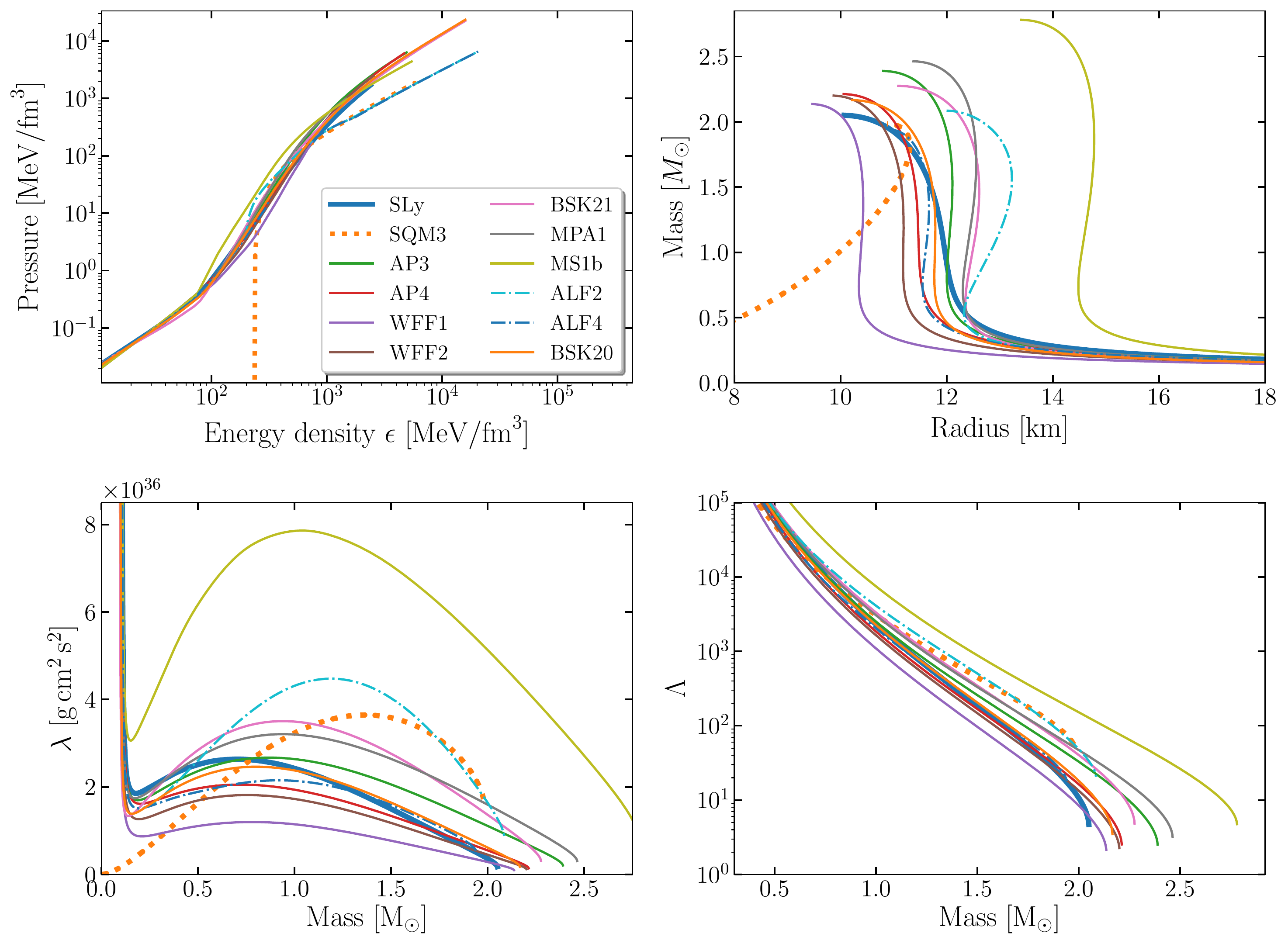}
\caption{\emph{Illustration of the differences on the physical parameters predicted by a representative sample of EoS.} Solid lines correspond to hadronic EoSs, dotted lines denote self-bound strange quark matter EoSs, and dot-dashed lines correspond to the hybrid EoSs containing mixtures of nucleonic and quark matter. EoSs SLy and SQM3 are highlighted by thicker lines here for comparison with Fig.~\ref{fig:4panelSLY}. \emph{Top Left:} Relation between pressure and energy density. Three EoS are particularly outstanding, corresponding to SQM3 and both ALF2 and ALF4, which are the only non-fully nucleonic EoS considered. In these examples, the non-nucleonic phases lead to smaller pressures and a ``softening" of the EoS. Here, units are $ c=1$. \emph{Top Right:} Mass versus radius relation as predicted by each EoS. It can be observed how the softer EoS containing non-nucleonic phases result in smaller maximum masses barely reaching 2$M_{\odot}$. The SQM3 EoS also shows a remarkable behaviour because the self-bound strange quark matter is incompressible, which results in the radius increasing with the mass, i.e. opposite trends than exhibited by hadronic EoSs. \emph{Bottom Left:} Tidal deformability as a function of mass. While most EoSs show a similar behaviour, MS1b stands out for its large deformability and radius for a given mass, and the SQM3 model again exhibits a different behavior because of the incompressibility of the matter. \emph{Bottom Right:} Dimensionless tidal deformability versus mass predicted by each EoS. Note that in this dimensionless  quantity, the SQM3 EoS does not show any exotic behaviour.}
\label{fig:4panelALLEOS}
\end{figure}

Linking these microscopic features to the macro-scale properties of a NS requires choosing an EoS, and solving equations~\eqref{backgroundTOV} and ~\eqref{eq:H0} for a range of central pressures. Figure~\ref{fig:4panelSLY} illustrates an example for the SLy~\cite{Douchin:2001sv} nucleonic EoS (solid blue curve) and the strange quark matter SQM3~\cite{Prakash:1996xs} EoS (dashed orange curve). Green squares and red diamonds represent equal central pressures for stars with these EoSs, leading to very different masses, radii and tidal parameters. Black triangles represent the maximum central pressure a star can support for that EoS. For SLy (and most EoSs, see Fig.~\ref{fig:4panelALLEOS}) a smaller central pressure yields a less massive NS, which is less compact and therefore, more deformable. Note, however, that these general trends do not hold for self-bound strange quark stars, where an increase in the mass leads to larger radii because the matter is incompressible, as represented by SQM3 in Fig.~\ref{fig:4panelSLY}. If the NS has a central pressure higher than the one corresponding to the maximum mass that this EoS can support, the NS will generally become unstable and collapse to a black hole. However, some classes of EoS models, e.g. those described in Ref.~\cite{Alford:2013aca} with strong phase transitions from hadronic to quark matter, give rise to multiple stable branches in the mass-radius plane for NSs. A perturbation may thus cause the NS to migrate to the other branch of solutions instead of forming a black hole.

While most EoSs are similar below $\rho_{sat}$, beyond that threshold, clear differences arise as seen in Fig.~\ref{fig:4panelALLEOS}. These differences then translate into very diverse mass to radius relations and tidal deformabilities, which are key observable quantities. Astrophysical constraints on mass and radius have been inferred from X-ray observations yet remain plagued by systematic uncertainties. However, highly anticipated new kinds of measurements of NS compactness-dependent gravitational lensing effects with NICER \cite{Watts:2019lbs} are about to be released. Importantly, the discovery of GW170817 has opened up the opportunity to use the tidal parameters to constrain the EoS, as considered here. 

The set of EoSs used for this analysis (obtained from the Xtreme catalogue~\cite{Ozel:2016oaf}, available at \cite{Xtreme}; for an alternative compilation of EoSs see e.g. the CompOSE library~\cite{Compose}) provides an illustrative but by no means complete sample of different EoS considered. While most of these EoSs consider only nucleonic matter, they differ in the treatment of the particle interactions. The SLy EoS \cite{Douchin:2001sv} is computed with a two potential-method, MPA1 \cite{Muther:1987xaa} employs a relativistic Brueckner-Hartree-Fock method, AP3-4 \cite{Akmal:1998cf} and WFF1-2 \cite{Wiringa:1988tp} use an eight variational-method, and lastly MS1b \cite{Mueller:1996pm} uses a three relativistic mean field method. Non-nucleonic matter EoSs were also considered, such as SQM3 \cite{Prakash:1996xs} as a self-bound strange quark matter model and ALF2-4 \cite{Alford:2004pf} with a mixture of nucleonic and quark matter. Only EoSs capable of yielding maximum masses above 1.95$M_{\odot}$ were considered in Fig.~\ref{fig:4panelALLEOS}, with the value chosen as an arbitrary cutoff close to the $\gtrsim 2M_\odot$ observational bound.

%%%%%%%%%%%%%%%%%%%%%%%%%%%%%%%%%%%%%%%%%%%%
\section{Tidal effects on the orbital dynamics and GWs }
\label{sec:dynamicsandGWs}
%%%%%%%%%%%%%%%%%%%%%%%%%%%%%%%%%%%%%%%%%%%%
We next consider the effect of the NS's tidal distortion on the orbital dynamics and GWs from a binary system. We restrict the discussion to the leading-order effects in the small parameters $\alpha=R/d\ll 1$, $\varepsilon_{\rm orbit}=GM/dc^2\sim v^2/c^2\ll 1$, and consider only the adiabatic quadrupolar tidal effects discussed above for nonspinning NSs on circular orbits. In this regime, the center-of-mass motion for point masses dominates the dynamics, and finite-size effects induce only small corrections as we briefly sketch below.  

The energy of the binary is the sum $E=E_{\rm point-mass}+E_{\rm tidal}$ with $E_{\rm tidal}={\cal Q}_{ij}{\cal E}^{ij}/4$ in the adiabatic case. For circular orbits, the leading-order contribution to $E_{\rm tidal}$ can be expressed explicitly in terms of the orbital variables and $\lambda$ as ~\cite{Vines:2011ud,Flanagan:2007ix,Hinderer:2009ca, Steinhoff:2016rfi, Wade:2014vqa}:
\be
\label{eq:tidalE}
E_{\rm tidal}= -\frac{1}{2}\mu c^2\left(\frac{GM_{\rm T}\Omega}{c^3}\right)^{2/3}\left[\frac{c^{10}}{G^4}\frac{\lambda_1}{M_{\rm T}^{5}}\left(-9\frac{M_2}{M_1} \right)\right]\left(\frac{GM_{\rm T}\Omega}{c^3}\right)^{10/3},
\ee
where $\Omega\sim \sqrt{GM/d^3}$ is the orbital angular frequency and $\mu=M_1M_2/M_{\rm T}$ is the reduced mass. The subscripts $1,2$ label the two objects in the binary. For two NSs, the contribution from the second NS adds linearly to all the expressions. 

The radiated GW power, to the leading order, is given by quadrupole formula $P^{\rm GW}=G/(5c^5)\langle \dddot{Q}^{\rm T}_{ij} \dddot{Q}^{\rm T\; ij}\rangle$. Here, overdots denote time derivatives and the angular brackets represent an average over several wavelengths of the GWs. The tensor $Q_{ij}^{\rm T}$ is the total quadrupole of the binary~\cite{Flanagan:2007ix} given by $Q_{ij}^{\rm T}=\mu d^2 (n^i n^j-\delta_{ij}/3)+Q_{ij}$, where the first term is due to the orbital motion of the NSs' centers of mass. The time-variation of $Q_{ij}$ is phase coherent with the orbital quadrupole since the NSs are not tidally locked, thus the tidal bulge pointing towards the companion travels around the NS surface during an orbit. By contrast, a tidally locked system such as the Moon has a spin period that is identical to its orbital period and thus always faces the Earth with the same side.

The result for the linearized tidal contribution to $P_{\rm GW}$ is~\cite{Flanagan:2007ix,Vines:2011ud}
\be
\label{eq:tidalflux}
P^{\rm GW}_{\rm tidal}=\frac{32c^5\mu^2}{5GM_{\rm T}^2} \left(\frac{GM_{\rm T}\Omega}{c^3}\right)^{10/3}\left[\frac{ c^{10} }{G^4}\frac{\lambda_1}{ M_{\rm T}^5}\left(\frac{18M_{\rm T}}{M_1}-12\right)\right]\left(\frac{GM_{\rm T}\Omega}{c^3}\right)^{10/3}.
\ee

An approximation for the chirp GW signal (of the form shown in Fig.~\ref{fig:waveforms}, sinusoidal with an increasing amplitude and frequency) measured by a distant observer can be obtained by imposing that on average, the GW power radiated by the binary system $P^{\rm GW}$ is compensated by a change in the binary's energy and angular momentum, leading to an adiabatic inspiral. This averaged energy balance implies that the orbital frequency $\Omega$ evolves according to $d\Omega/dt=-P^{\rm GW}/(dE/d\Omega)$. The GWs in this approximation oscillate at twice the orbital frequency $f=\Omega/\pi$, where $f$ is the GW frequency. For data analysis, it is useful to compute the Fourier transform of the GW signal which is approximately given by~\cite{Cutler:1994ys} 
\be
\label{eq:tildeh}
\tilde{h}(f)={\cal A}f^{-7/6}\exp\left[i\left(\psi_{\rm point-mass}+\psi_{\rm tidal}\right)\right].
\ee
Here, ${\cal A}$ is an amplitude (including both point-mass and matter effects) and $\psi$ denotes the phase, which the GW measurements are especially sensitive to~\cite{Cutler:1992tc, Cutler:1994ys}. The tidal corrections to the phase $\psi_{\rm tidal}$ can be computed by solving the energy balance relation in the form
\be
\frac{d^2\psi}{df^2}=-2\pi^2 \frac{(dE/d\Omega)}{P^{\rm GW}}. 
\label{eq:ddpsi}
\ee
Using the results for $E$ and $P^{\rm GW}$ with the tidal contributions from Eqs.~\eqref{eq:tidalE} and \eqref{eq:tidalflux}, and linearizing in the tidal corrections leads to the following explicit result for the tidal signature in the frequency-domain GW phasing~\cite{Flanagan:2007ix}
\bes
\be\label{eq:deltapsi}
\psi_{\rm tidal}=\frac{3}{128(\pi G{\cal M} f/c^3)^{5/3}}\left[ -\frac{39}{2}\tilde{\Lambda}\, (\pi G M_{\rm T} f/c^3)^{10/3} \right].
\ee
The prefactor here is the leading order result for point masses involving the chirp mass, 
which is a combination of the individual masses ${\cal M}=(m_1 m_2)^{3/5}/(m_1+m_2)^{1/5}$. The chirp mass represents the best-measured GW parameter for long inspiral signals such as from binary NSs. Similarly, the leading-order tidal GW signature is characterized by a weighted average combination of the individual contributions $\tilde \Lambda$ given by
\be
\label{eq:lambdatilde}
\tilde{\Lambda}=\frac{16c^{10}}{13G^4M_{\rm T}^5}\left[\left(1+\frac{12 M_2}{M_1}\right) \lambda_1+\left(1+\frac{12 M_1}{M_2}\right)\lambda_2\right].
\ee
\ees
Analogous to the chirp mass, this combination is the best-measured GW parameter characterizing tidal effects for inspiral signals. This particular combination~\eqref{eq:lambdatilde} arises from the different dependencies of $E_{\rm tidal}$ and $P^{\rm GW}_{\rm tidal}$ on the masses (c.f. Eqs.~\eqref{eq:tidalE} and ~\eqref{eq:tidalflux}) which linearly combine when perturbatively expanding the right hand side of Eq.~\eqref{eq:ddpsi} for small tidal corrections to compute the effect on the phase. For equal-mass binary NSs $\tilde{\Lambda}$ reduces to $\Lambda$ of the individual NSs. 

For practical data analysis more sophisticated models of tidal effects are used, e.g. those of Refs.~ \cite{Buonanno:1998gg,Buonanno:2000ef, Ajith:2007qp, Ajith:2007kx,Bohe:2016gbl,Babak:2016tgq,Nagar:2017jdw,Khan:2015jqa,Schmidt:2014iyl,Dietrich:2017aum,Bernuzzi:2014owa,Hinderer:2016eia,Steinhoff:2016rfi}. We note that tidal effects in the phase scale as a high power of the GW frequency relative to the leading order point-mass result outside the brackets in Eq.~\eqref{eq:deltapsi}. This implies that they become most important in the later parts of the inspiral. Furthermore, when considering the scaling with the frequency the tidal corrections in the phase appear effectively as would a high-order post-Newtonian contribution. However, they are in fact the leading-order terms in the finite-size expansion for nonspinning NSs within the double expansion scheme discussed in Sec.~\ref{sec:theory}, where $\alpha$ and $\epsilon_{\rm orbit}$ depend on the orbital separation and hence frequency in the same way.

The GW parameter estimation yields multiple source parameters. We focus here on the chirp mass $\cal M$, the mass ratio $M_2/M_1$, and $\tilde{\Lambda}$. The relation between $\tilde \Lambda$ and the mass parameters is EoS dependent, and thus, can be used to constrain the EoS. We present their relation in Fig.\ref{fig:2DLTILDEMchirp}, where four distinct EoSs are considered. In order to obtain these predictions, we explored the range of possible $\cal M$ values for NS masses with $M_{\text{min}}=1M\odot$ and $M_{\text{max}}$ given by the EoS considered. Figure~\ref{fig:2DLTILDEMchirp}  shows that large values of $\cal M$ relate to a very narrow range of values for $\tilde{\Lambda}$ for most EoSs. Hence, GW measurements with large values of $\cal M$ will prompt sharper EoS predictions, which can be utilized to constrain the EoS.

\subsection{Example application to GW170817}

The GW discovery of the binary NS inspiral event GW170817 has spurred a plethora of analyses and EoS constraints by many different groups, some of which also included information from the electromagnetic counterparts; see e.g. the over two thousand citations to the LIGO-Virgo Collaboration papers~\cite{TheLIGOScientific:2017qsa, Abbott:2018wiz, Abbott:2018exr}. Here, we will restrict the discussion to a simple illustration of the basic results obtained from the GW side as described in detail in the source properties paper by the LIGO Virgo collaboration \cite{Abbott:2018wiz} and for which results of the Bayesian parameter inference are publicly available \cite{PropertiesDATA}. The reader should also refer to \cite{Abbott:2018wiz} for more detailed analyses. The results given in \cite{Abbott:2018wiz,PropertiesDATA} are obtained with two different choices of priors on the spins (high and low spin priors), assuming a flat prior on $\tilde{\Lambda}$, and obtained with the waveform model known as PhenomPNRTidal~\cite{Hannam:2013oca,Schmidt:2012rh, Schmidt:2014iyl, Dietrich:2017aum}. While other waveform models exist and were used in \cite{Abbott:2018wiz} to demonstrate that for GW170817 the statistical errors dominated over the quantifyable systematic uncertainties due to modeling and detector calibration, the posteriors are not publicly available and could thus not be included here. 
 
For GW170817, the observed GW signal was from the binary inspiral epoch. As discussed in Sec.~\ref{sec:dynamicsandGWs} for such cases the chirp mass is the best measured parameter, and for GW170817 was determined to be ${\cal M}_{\rm GW170817}=1.186_{-0.001}^{+0.001}M_{\odot}$. However, the mass ratio $M_2/M_1$ -which in combination with ${\cal M}$ would yield the individual masses- was less well measured. The fascinating aspect of the GW measurements of GW170817 was that for the first time the tidal deformability parameter $\Lambda_{1,2}$ of the objects and hence the EoS could be constrained from GW data. Additional matter effects, for instance from a rotationally-induced quadrupole in spinning NSs appeared to be too small to measure, as far as could be assessed in the analysis of Ref.~\cite{Abbott:2018wiz}, and effects due to possible nonlinear tidal instabilities~\cite{Xu:2017hqo} and from the postmerger signals ~\cite{Abbott:2018wiz} were also indiscernible.

%last fig used to be here

\begin{figure}[h]
\centering
\includegraphics[width=0.7\textwidth]{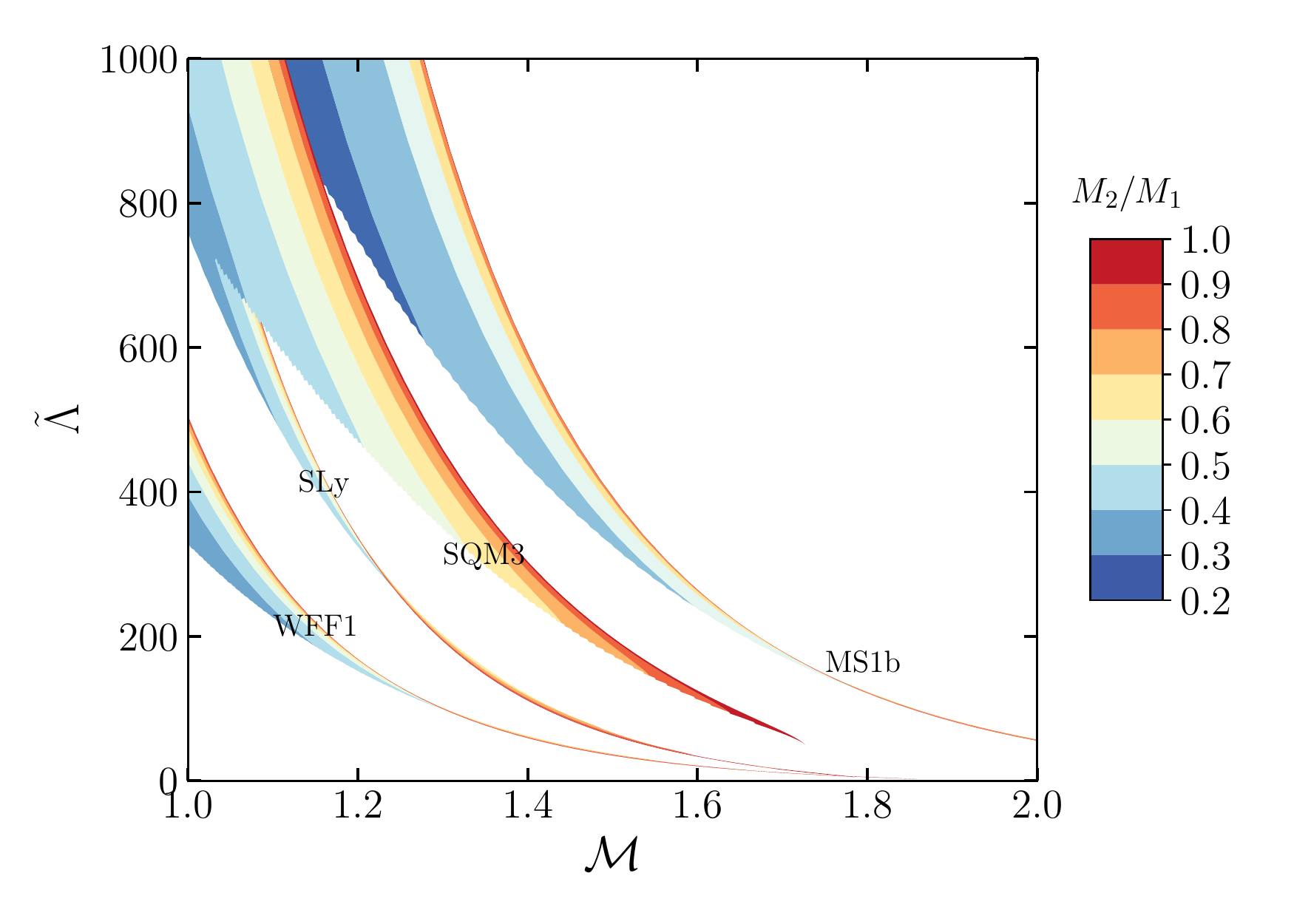}
\caption{ Relation between the dimensionless $\tilde{\Lambda}$, $\cal M$ and the mass ratio.
We considered four distinct EoSs aiming to represent the wide range of EoS candidates, from the stiffer (MS1b) to the softer (WFF1). For each EoS, larger values of $\cal M$ correlate with narrower ranges of allowed $\tilde{\Lambda}$. We also observe that small values of $\tilde{\Lambda}$ can only be encountered for nearly equal masses. We also note that for a fixed ${\cal M}$ and EoS an equal-mass binary yields the largest values of $\tilde \Lambda$. } 
\label{fig:2DLTILDEMchirp}
\end{figure}

\begin{figure}
    \includegraphics[width=0.49\textwidth]{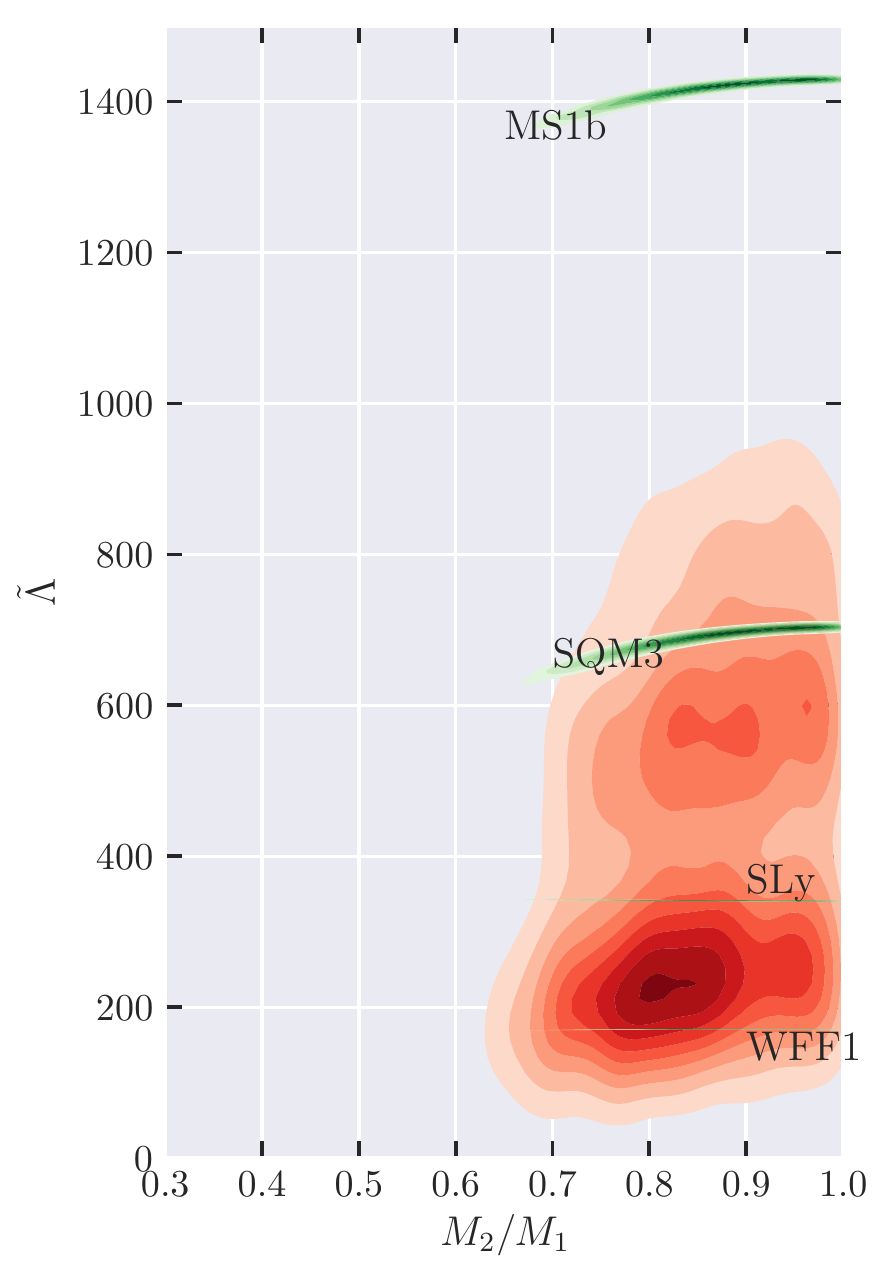}
    \includegraphics[width=0.49\textwidth]{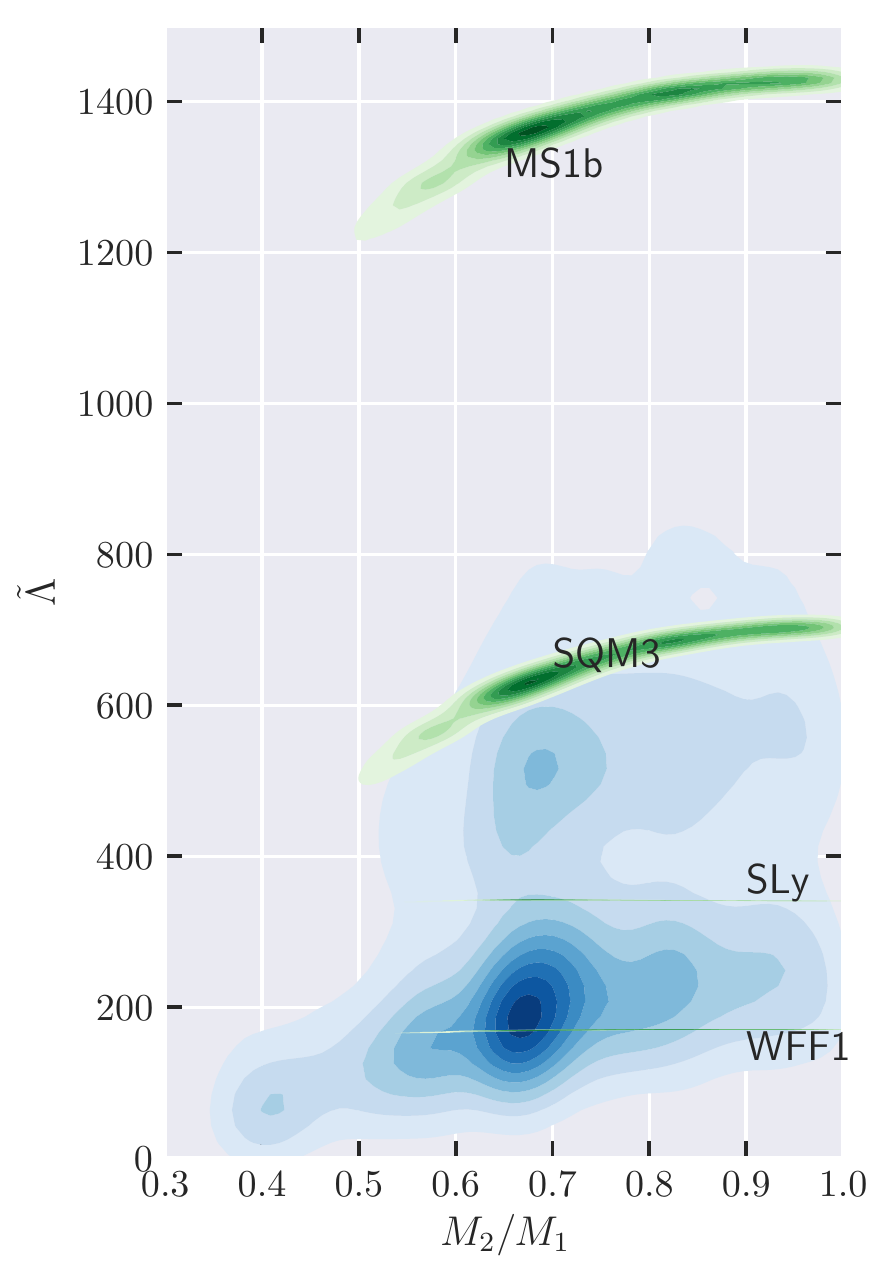}
    \caption{Posterior density functions from the Bayesian analysis using the PhenomPNRTidal model provided by the LIGO Virgo collaboration \cite{PropertiesDATA}. The two panels correspond to different prior assumptions on the spins of the NSs: a low spin prior (left panel) and a high spin prior (right panel). Green-colored regions illustrate the tidal parameters predicted by four EoSs using the masses from the corresponding posterior samples. For the EoSs SLy and WFF1 these contours are extremely sharp and hardly visible, this is a consequence of their tight ${\cal M}$-$\tilde{\Lambda}$ relation for large ${\cal M}$ values. Even though large uncertainties remain as the credible region is large, very stiff EoSs candidates such as MS1b are disfavoured. The relatively straight lines of the EoSs in the $\tilde\Lambda$-mass ratio plane result from the proportionality relation between $\tilde{\Lambda}$ and the mass ratio for a fixed ${\cal M}$ illustrated in Fig.\ref{fig:2DLTILDEMchirp}. This relation is a direct consequence of the range of allowed mass ratios each EoS supports for a given $\cal M$. This can be observed in Fig.\ref{fig:2DLTILDEMchirp}, where a vertical line crossing each EoS would contain a range of allowed mass ratios, as observed through the contours.}
    \label{fig:LTILDEvsQpdfLS}
\end{figure}

The source parameters are inferred from GW data by employing a Bayesian analysis, as executed by \cite{Abbott:2018wiz}. The posteriors from GW170817 for the PhenomPNRTidal are publicly available at \cite{PropertiesDATA}. In Fig.~\ref{fig:LTILDEvsQpdfLS} we plot the posterior densities for $\tilde{\Lambda}$ and mass ratio $M_2/M_1$ for the low and high spin priors computed from the data for the masses and individual tidal parameters available in \cite{PropertiesDATA}. Additionally, we calculated the tidal parameter predicted by a subset of EoS using the masses corresponding to each sample. The EoS subset employed here is the same as for Fig.\ref{fig:2DLTILDEMchirp}, representing a wide range of EoS models that have been proposed. Both priors overlap in the region of mass ratios $M_2/M_1\leq 0.7$, however, the results with the high spin prior support a larger range of allowed mass ratios because spins and mass ratio are to some extent correlated in GW measurements.  
While very stiff EoSs such as MS1b are notably far from the probability intervals and hence disfavored by the measurements (they are also disfavored by nuclear physics constraints but included here for illustration), the overall constraints remained relatively weak. Using the full information on individual tidal deformabilities in the waveform, assumptions that the event was a NS-NS binary with small spins, and the multi-wavelength observations of the electromagnetic counterpart have all be used in a variety of studies to further tighten the EoS constraints inferred from this event. However, because of the additional assumptions required and the large uncertainties in modeling the multimessenger counterparts we do not consider this additional information here in the context of reviewing implications from GW170817. Progress on modeling is already underway, and will likely enable more robust multimessenger EoS constraints in the future. 

\subsection{Summary and outlook} 

%--------------------------------------------------------------------------------
We have reviewed here the basic physics, calculations, and GW signatures associated with adiabatic quadrupolar tidal effects that occur during a binary inspiral and are predominantly characterized by the EoS-dependent tidal deformability parameter. These effects are of key interest for using GWs as a novel tool for probing the EoS of matter in NSs, as was initiated with the measurements of the GW170817. While the analysis of GW170817 already led to interesting constraints, see e.g.~\cite{Abbott:2018exr} and numerous other papers following up on GW170817 (see citations to the discovery paper~\cite{TheLIGOScientific:2017qsa}), the main impact of GWs on advances in subatomic physics will come with observations in the near future, as the GW detectors increase in their sensitivity and observe populations of NS mergers. This will enable more precise measurements of $\tilde\Lambda$ and mapping out the $\Lambda$ versus mass relation with multiple events that will likely differ in the constituent masses. Additional insights on NS matter will come from the highly anticipated discoveries of NS-black hole binaries where the possible tidal disruption of the NS could be measurable as a characteristic shutoff in the GW signal~\cite{Vallisneri:1999nq,Shibata:2011jka,Ferrari:2009bw,Maselli:2013rza,Foucart:2014nda, Foucart:2012vn,Kawaguchi:2017twr,Lackey:2013axa}, see the review article on NS-BH mergers~\cite{Shibata:2011jka} for details. Furthermore, as also exemplified by the discoveries and analyses of GW170817, the electromagnetic counterparts yield complementary invaluable insights and constraints on the EoS and nuclear matter, see e.g. the review articles~\cite{Metzger:2016pju,Tanaka:2016sbx}.

At higher sensitivity, the GW measurements with current facilities may also discern other matter effects during the inspiral or even detect the strongest portion of the postmerger signals. Additional matter effects during the inspiral include phenomena that contain similar EoS information to that encapsulated in $\Lambda$, with the corresponding characteristic parameters being related to $\Lambda$ in a nearly EoS-independent way, at least for a wide range of proposed EoS models within GR~\cite{Yagi:2014qua, Yagi:2013sva, Yagi:2013baa, Yagi:2013awa}. Examples of such effects include rotational multipole moments~\cite{Laarakkers:1997hb, Poisson:1997ha,Harry:2018hke,Krishnendu:2017shb}, higher tidally induced multipole moments~\cite{Yagi:2016qmr}, and the frequency-dependent response due to the fundamental modes~\cite{Steinhoff:2016rfi, Hinderer:2016eia, Schmidt:2019wrl, Pratten:2019sed,Chan:2014kua, Chirenti:2016xys, Gold:2011df} when relaxing the restrictive specialization to the adiabatic limit in Sec.~\ref{sec:definitionoflambda}. The frequency of the tidal disruption GW signature in NS-black hole binaries~\cite{Pannarale:2015jia} and the main feature of the NS-NS postmerger signals arising from the rotation of the remnant in cases without prompt BH formation also exhibit quasi-universal relations of their associated dominant characteristic parameters to $\Lambda$; see e.g. references in the review~\cite{Hinderer:2018mrj}.

In the longer term, potential future instrumental upgrades to existing facilities such as the A+ detectors~\cite{Miller:2014kma, APlusDesign} or targeted improved high-frequency sensitivity~\cite{Miao:2017qot,Martynov:2019gvu}, and the ambitious third-generation instruments such as the Einstein Telescope~\cite{Punturo:2010zz, Hild:2010id} and the Cosmic Explorer (CE) observatories~\cite{Evans:2016mbw}
will provide unprecedented access to complementary information on further details of matter in NS interiors, even during the inspiral regime. This could be achieved by observing a variety of other oscillation modes associated with details of the composition such as phase transitions to hyperonic or quark matter~\cite{Ho:1998hq, Flanagan:2006sb, Shibata:1993qc, Yu:2016ltf, Lai:1993di, Kokkotas:1995xe,Tsang:2013mca, Tsang:2011ad}, measuring other kinds of tidal effects that arise in GR known as gravito-magnetic tides~\cite{Flanagan:2006sb, Pani:2018inf}, discerning spin-tidal couplings~\cite{Jimenez-Forteza:2018buh}, and potential nonlinear tidal instabilities~\cite{Xu:2017hqo,Essick:2016tkn, Landry:2015snx}. These future instruments will also have the exciting capability to measure details of the postmerger spectrum, thus probing the EoS in completely unexplored regimes at higher densities and temperatures than attained in NSs during the inspiral or in isolation, and yielding insights into the microphysics of the merger and black hole formation.

We emphasize that our discussion of the physics during the inspiral was limited to basic considerations. Corrections from other effects are expected to become important especially in the later parts of the inspiral, when the approximations we have outlined start to break down. Recent theoretical developments have addressed some of these issues, and also profited from advances in numerical relativity simulations that serve as important tests and in some cases inputs for the tidal models used in GW data analysis. Finally, we have also omitted any discussion of finite-size effects in alternative theories of gravity, theories of physics beyond the standard model, exotic objects, and bound states of fundamental fields, all of which are interesting applications described e.g. in Ref.~\cite{Barack:2018yly}. In conclusion, the first constraints on tidal parameters by LIGO and Virgo with GW170817~\cite{TheLIGOScientific:2017qsa} have made GWs available as a new probe of the nuclear equation of state, in conjunction with the additional information contained in the multimessenger counterparts. However, this event marked only the beginning of using GWs as a new tool for probing subatomic physics. We anticipate an increasing wealth of new insights from the deeper explorations and studies enabled by future observations that will accumulate over the coming years.

\section*{Acknowledgements}

\noindent We thank Geert Raaijmakers for useful discussions and help with the numerical implementation. AGC thanks the GRAPPA institute at the University of Amsterdam and Samaya Nissanke's research group for hospitality during his bachelor's
project. TH gratefully acknowledges support from NWO Projecruimte grant GW-EM NS and the DeltaITP. 

%\section{References}
\bibliographystyle{unsrt.bst}
\bibliography{tidal}
\end{document}